\documentclass[
 reprint,
 amsmath,amssymb,
 aps,
]{revtex4-2}

\usepackage{graphicx}
\usepackage{bm}
\usepackage{lineno}
\usepackage{siunitx}
\usepackage{url}
\usepackage{hyperref}

\usepackage{dcolumn}
\newcolumntype{d}[1]{D{.}{.}{#1}}

\usepackage[ruled,vlined]{algorithm2e}
\usepackage{algorithmic}
\begin{document}

\title{Universal and Experiment-calibrated Prediction of XANES through Crystal Graph Neural Network and Transfer Learning Strategy}

\author{Zichang Lin, Wenjie Chen, Yitao Lin, Xinxin Zhang, Yuegang Zhang}
\affiliation{
 Department of Physics, Tsinghua University, Beijing 100084, China
}

\begin{abstract}
Theoretical simulation is helpful for accurate interpretation of experimental X-ray absorption near-edge structure (XANES) spectra that contain rich atomic and electronic structure information of materials. However, current simulation methods are usually too complex to give the needed accuracy and timeliness when a large amount of data need to be analyzed, such as for in-situ characterization of battery materials. To address these problems, artificial intelligence (AI) models have been developed for XANES prediction. However, instead of using experimental XANES data, the existing models are trained using simulated data, resulting in significant discrepancies between the predicted and experimental spectra. Also, the universality across different elements has not been well studied for such models. In this work, we firstly establish a crystal graph neural network, pre-trained on simulated XANES data covering 48 elements, to achieve universal XANES prediction with a low average relative square error of 0.020223; and then utilize transfer learning to calibrate the model using a small experimental XANES dataset. After calibration, the edge energy misalignment error of the predicted S, Ti and Fe K edge XANES is significantly reduced by about 80\%. The method demonstrated in this work opens up a new way to achieve fast, universal, and experiment-calibrated XANES prediction.
\end{abstract}

\maketitle

\section{Introduction}

X-ray absorption spectroscopy (XAS) is a powerful tool to detect the electronic structures and atomic geometry of materials, including atom valence, bond length and coordination number\cite{RN150}. It has the advantages of non-destructive, element selectivity, and high sensitivity\cite{RN198}. In addition, XAS is available for various sample types, such as bulk materials, powders, and solutions\cite{RN315}. Therefore, it has been widely applied to study the problems in different areas, for example, low dimensional materials characterization\cite{RN317,RN318}, catalytic mechanism analysis\cite{RN244,RN195}, relics identification\cite{RN316}, and biomedical analyses\cite{RN241}. Especially, these unique advantages enable XAS to combine with the operando/in-situ techniques for real-time characterization of some complex systems, like electrochemistry reactions in batteries\cite{RN134,RN133,RN174,RN358}. According to the energy region with respect of the absorption edge, XAS can be divided into X-ray absorption near-edge structure (XANES) and extended X-ray absorption spectroscopy (EXAFS). EXAFS is mainly dominated by single scattering events, where coordination numbers and interatomic distances can be derived from Fourier transform for some materials with periodic structure\cite{RN176,RN124}. XANES is related to strong scattering processes as well as local atomic resonances, which is more complicated than EXAFS and hard to directly transformed to structure parameters\cite{RN230}. Commonly, interpretation of experimental XANES is based on some empirical features, such as comparing edge shifts and peak positions with referenced spectra of standard materials\cite{RN176}. 

With the development of XAS theory and the corresponding calculation software like FEFF9\cite{RN217}, FDMNES\cite{Bunău_2009}, and OCEAN\cite{RN251}, the XANES can be quantitively simulated from a given crystal or molecule structure and taken as the referenced spectra in analysis. It helps to gain deeper insight into the relationship between local environment of absorbing atoms with spectra features. However, such calculation is time-consuming when the materials structure is complex. And the calculated spectra are sensitive to the selected approximation methods and calculation parameters, where a tedious parameters optimization process is needed to obtain reasonable results. In addition, XANES simulation faces challenges in aligning absolute energy scales and accounting for experimental broadening effects, resulting in inevitable discrepancies compared with experiments\cite{RN249}. Without careful experimental calibration, the simulated XANES may yield false edge and peak features, misleading the XANES analysis. These shortcomings of current simulation methods have brought significant challenges for efficient and accurate XANES interpretation. This especially limits the in-situ XANES study of some essential but complex systems requiring massive data analysis, such as battery materials that experience constant structure changes during charge and discharge processes.  

The development of artificial intelligent (AI) has opened up a new paradigm to solve the similar complex problems in different areas, such as AlphaFold for protein structure prediction\cite{RN253}, GNoME for material stable structures discovery\cite{RN256}, and DeepH for Hamiltonian calculation in DFT\cite{RN255}. Researchers have also made some attempts to apply AI methods in XANES prediction to avoid the complex calculation process. These developed models cover the predictions on C K edge XANES of amorphous carbon and organic molecules\cite{RN121,RN142}, O and N K edge XANES of molecules\cite{RN138}, and 3d transition metals K edge XANES of transition metal compounds\cite{RN365,RN192,RN367}. Although those models offer a more efficient and convenient way to obtain XANES, their universality is challenged for they only focus on limited range of absorption elements. The universality of AI based XANES prediction on other common elements need to be further studied. In addition, due to the shortage of experimental spectra, the existing models are trained based on the high throughput simulated XANES, where the spectra features and energy scale are not accurately calibrated. The uncorrected discrepancies between the training data and experiment measurement challenge the reliability of existing XANES prediction models. Therefore, the universality and fidelity of AI based XANES prediction should be improved before it can be put into practical use. 

\begin{figure*}[t]
\includegraphics[width=0.6\textwidth]{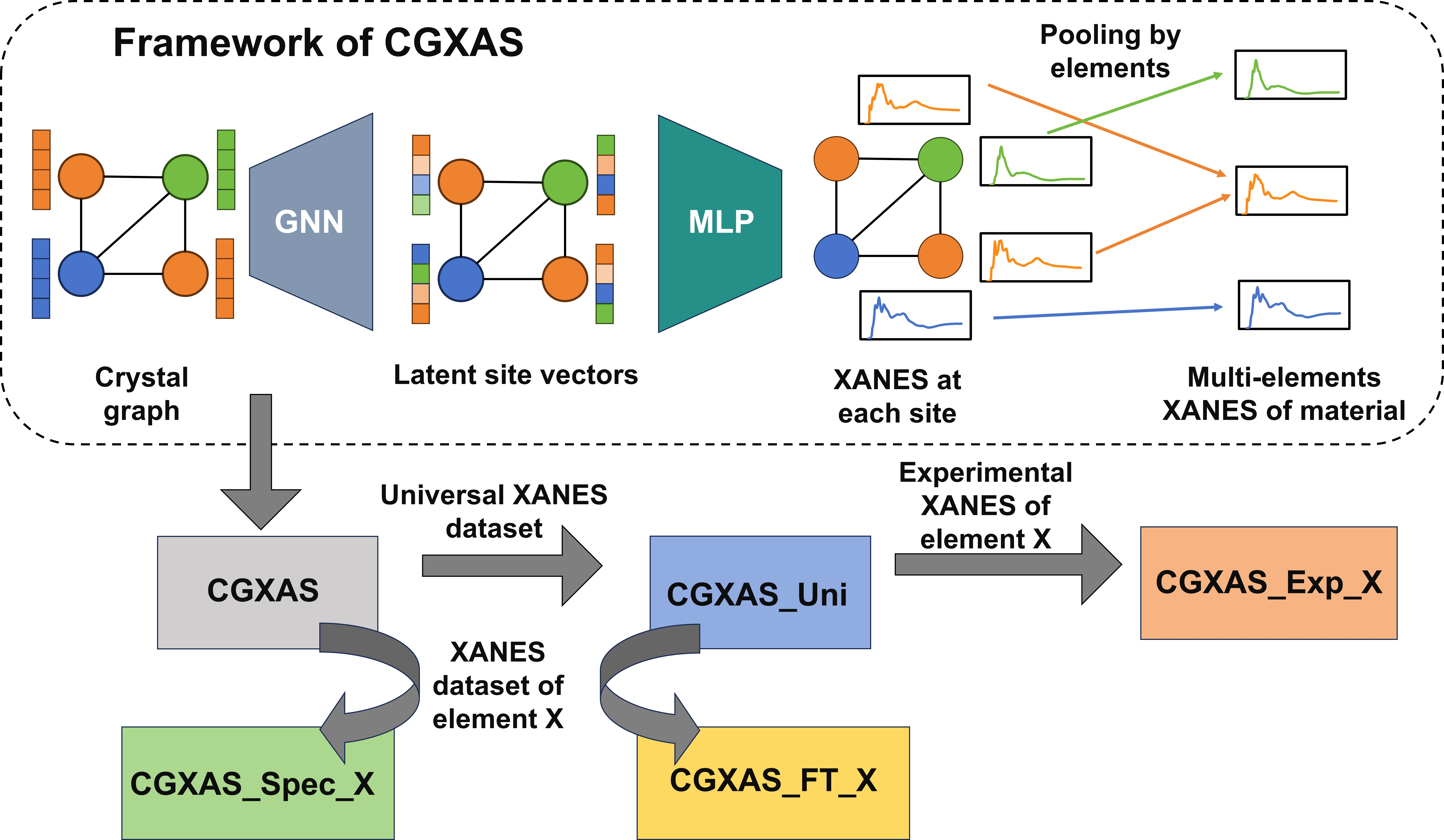}
\caption{\label{fig:workflow} Framework of CGXAS model and flowchart of training.}
\end{figure*}

In this work, we explored the universal and experiment-calibrated prediction of XANES through crystal graph neural network and transfer learning strategy. A crystal graph neural network model for XANES prediction (CGXAS) is firstly established, and the universality of CGXAS on common detected elements of K edge XANES is studied. Then, transfer learning strategy is adopted to calibrate the predicted XANES using a small experimental dataset. The framework of CGXAS and the training flowchart is represented in Fig.\ref{fig:workflow}. In CGXAS, the graph neural network (GNN) encodes the local environment information into a latent site vector for each atom in the crystal graph. Then, the multilayer perceptron (MLP) decodes the site vector into the corresponded K edge XANES of each site. The successive mask pooling layer is used to average the XANES at each site according to the element, enabling XANES prediction for different elements in one network. A universal dataset with 341405 simulated XANES covering 48 elements\cite{RN219} is used for CGXAS training and evaluation. This study compares three distinct approaches for achieving universal XANES prediction: (1) Training a single model (CGXAS\_Uni) on the full dataset for universal XANES prediction; (2) Training individual element-specific CGXAS models using the element-specific subsets and assembling them for universal XANES prediction, where model for element X is denoted as CGXAS\_Spec\_X; (3) Fine-tuning CGXAS\_Uni model on each element-specific subset, leading to CGXAS\_FT\_X model that represents the fine-tuned version for element X. Among the three models, the single CGXAS\_Uni model demonstrated the best overall accuracy and universality. To study the interpretability of CGXAS, the t-SNE method is utilized to analyze the latent site vectors. Furthermore, to bridge the gap between simulation and experiment, transfer learning is also adopted to adapt the pretrained CGXAS\_Uni model with a small experimental XANES dataset of element X, where the model after transfer learning is denoted as CGXAS\_Exp\_X. Compared with traditional learning process, transfer learning can transfer the knowledge learned in other domains to enhance the prediction performance on target domain\cite{RN319}. Since the prediction performance of neural network is closely related with training data size, transferring simulated XANES knowledge to experimental XANES prediction can compensate the insufficiency in experimental data quantity. Here, we demonstrate the ability of transfer learning to calibrate the XANES prediction of S, Ti and Fe K edge XANES using only 48, 40 and 45 experiment spectra for each element, where S, Ti, and Fe are the critical elements in the study of energy and catalysis. The calibrated CGXAS\_Exp\_S, CGXAS\_Exp\_Ti, and CGXAS\_Exp\_Fe models reduce the edge energy misalignment error of the predicted XANES by about 80\%.

\section{Results}
\subsection{Universal XANES Prediction Performance of CGXAS}
\begin{figure}[ht]
\includegraphics[width=\columnwidth]{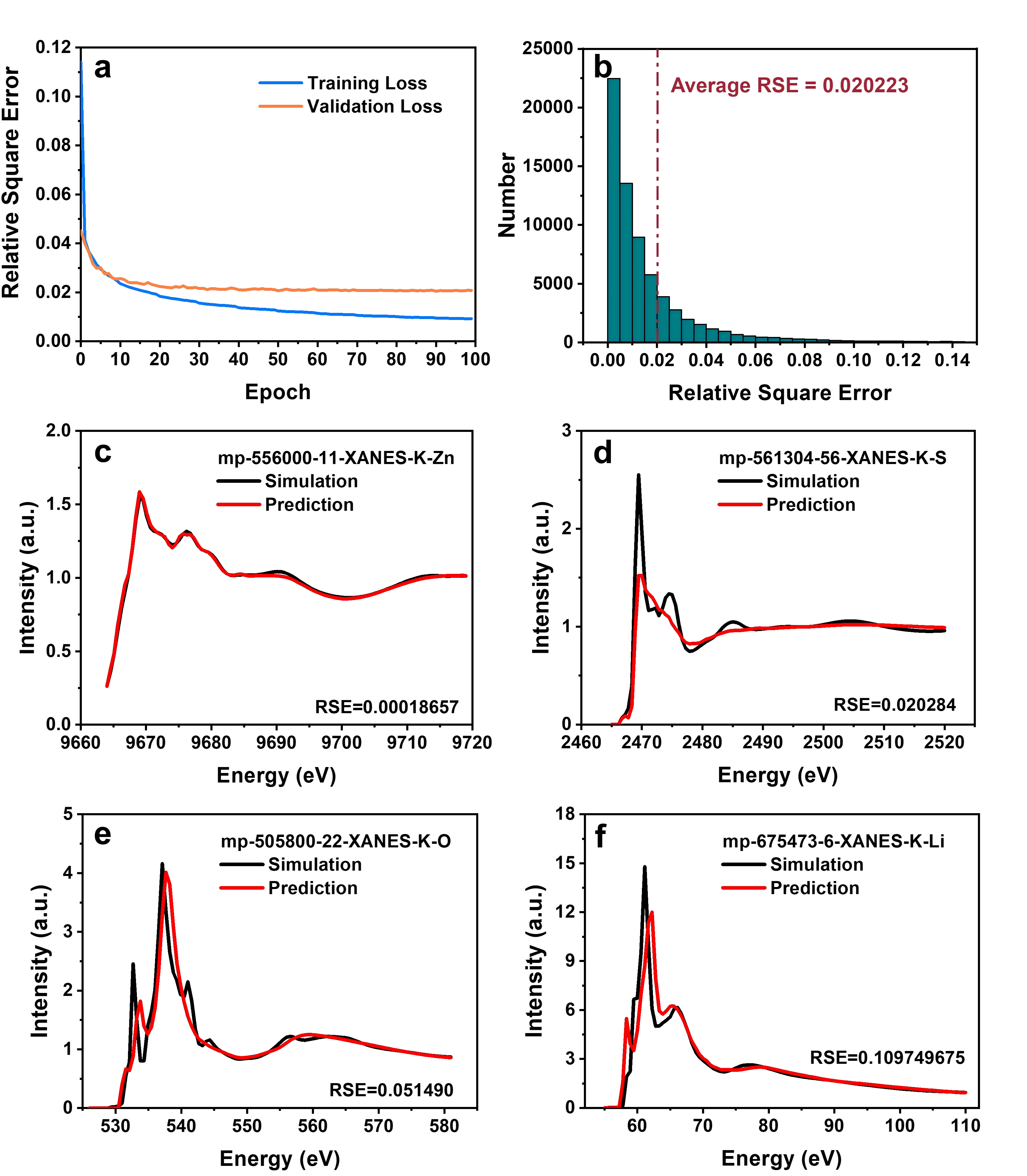}
\caption{\label{fig:train_val} Performance of CGXAS\_Uni on the prediction of XANES. (a) Evolution of training loss and validation loss at each epoch in the training process, where the loss is measured by relative square error (RSE). (b) Distribution of prediction error in the testing set. (c)-(f) Examples of the predicted XANES with (c) low, (d) median, (e) high, and (f) ultra-high prediction error compared with simulated XANES.}
\end{figure}

CGXAS\_Uni is trained on the universal dataset which contains the K edge XANES spectra of elements in period 2 to 5 (noble gas elements are excluded). Fig.\ref{fig:train_val}a shows the evolution of training loss and validation loss, represented by the relative square error (RSE), in the 100 epochs. The definition of RSE is shown in Eq.\ref{eq:RSE} at Methods section, where a RSE value closer to 0 stands for lower prediction loss. The training loss is reduced from 0.11 to 0.0093 after 100 epochs. And the validation loss converges to around 0.0207 after 70 epochs, which shows that 100 epochs are enough for model convergence. The distribution of test loss is shown in Fig.\ref{fig:train_val}(b), where most of the samples have the loss lower than 0.04. The average test loss is 0.020223, which is close to the validation loss. Examples of the predicted XANES spectra with low, medium, high, and ultra-high loss are shown in Fig.\ref{fig:train_val}(c)-(f), together with their simulated XANES spectra. It can be seen that CGXAS\_Uni can predict the general spectrum features of edge energy and main peak reasonably well, and the prediction errors mainly come from the differences in peak intensity, peak width, and peak energy. 

Besides training a single universal model, universal prediction of XANES can also be realized by assembling the element-specific models. The element-specific models can be either individually trained (CGXAS\_Spec\_X) or fine-tuned from CGXAS\_Uni model (CGXAS\_FT\_X) on the element-specific subsets of universal dataset. To compare the performances of these distinct approaches for universal XANES prediction, the prediction accuracy of CGXAS\_Uni, CGXAS\_Spec\_X, and CGXAS\_FT\_X on each absorption element is measured by the Pearson correlation coefficient (PCC) in testing set and shown in Fig.\ref{fig:cgxasuni}(a). Here, a PCC value closer to 1 represents higher accuracy in prediction. For most of the elements, their CGXAS\_Spec\_X models can have high prediction accuracy (PCC$>$0.88). But for Be, Mg, Cd, Sb, and I, their CGXAS\_Spec\_X models have relative low accuracies (PCC$<$0.83). Compared with CGXAS\_Spec\_X models, CGXAS\_Uni model yields enhanced prediction accuracies in almost all the elements. Especially for Be, Cd, Sb, and I, their PCCs are above 0.85 in CGXAS\_Uni model; and O, S, Cl, and Se are the only 4 elements whose PCCs in CGXAS\_Uni model are lower but with a negligible difference ($<$0.001) with those in CGXAS\_Spec\_X model. In general, therefore, the CGXAS\_Uni model achieves better performance in universal XANES prediction than CGXAS\_Spec\_X models. The CGXAS\_FT\_X models have similar performances in all elements compared with CGXAS\_Uni model. The average difference of PCC before and after fine-tuning is only around 0.002, showing that fine-tuning CGXAS\_Uni on element-specific dataset has little influence on its accuracy. Comparing the XANES prediction performances of CGXAS\_Uni, CGXAS\_Spec\_X and CGXAS\_FT\_X, it can be concluded that both accuracy and universality can be achieved by using the single CGXAS\_Uni model. Therefore, the following interpretability and transfer learning study will only be conducted based on the CGXAS\_Uni model. 

\subsection{Interpretability of CGXAS in XANES Prediction}
\begin{figure*}[ht]
\includegraphics[width=0.8\textwidth]{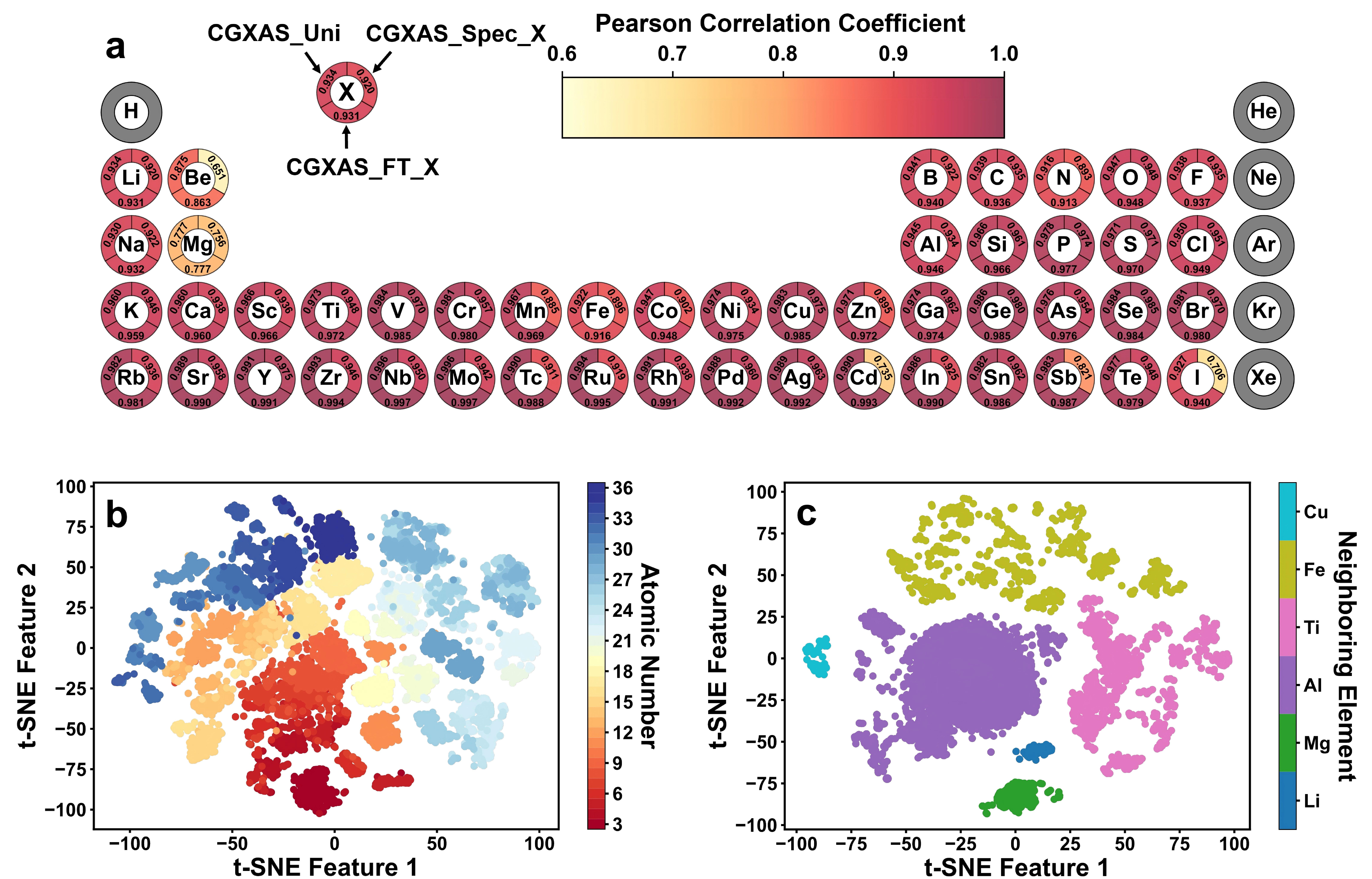}
\caption{\label{fig:cgxasuni} Universality and interpretability analysis of CGXAS. (a) Comparison of XANES prediction accuracy (measured by Pearson correlation coefficient) of CGXAS\_Uni, CGXAS\_Spec\_X, and CGXAS\_FT\_X in different elements. (b) Distribution of latent site vectors of different absorption sites, where the site vectors are reduced into 2 t-SNE features and colored according to their absorption elements. (c) Distribution of O latent site vectors obtained from Li, Mg, Al, Ti, Fe, and Cu metal oxides, where the site vectors are reduced into 2 t-SNE features and colored according to their neighboring elements. The latent site vectors in (b) and (c) are derived from CGXAS\_Uni model.}
\end{figure*}

XANES is determined by both the absorption atom and its local atomic environment\cite{RN230}. To analyze the interpretability of CGXAS on XANES prediction, it is essential to verify whether the features of absorption atom and local atomic environment can be identified from the latent site vectors after the information gathering by graph neural network. Here, the t-distributed stochastic neighbor embedding (t-SNE) method\cite{RN323} is applied to reduce the high dimension latent site vectors into 2 t-SNE features, so that the they can be plotted in 2D space. Fig.\ref{fig:cgxasuni}(b) shows the distribution of latent site vectors obtained from different absorption sites of various materials in the space of t-SNE features. Here, each element is marked by one specific color, which can be referred by its atomic number in the color bar. In the figure, the points with the same color tend to form a cluster, and clusters of different colors are separated obviously. It means that the information of absorption element is still kept in the latent site vector, which can be extracted by the t-SNE method. 

The t-SNE features distribution of O latent site vectors obtained from oxides of different metal elements (Li, Mg, Al, Ti, Fe, and Cu) is shown in Fig.\ref{fig:cgxasuni}(c), where the points of O sites are colored according to their neighboring elements. In the t-SNE feature space, these points are grouped by the neighboring elements, and different groups are clearly separated. It indicates that the information of local atomic environment can also be extracted from the latent site vectors.

\begin{figure*}[t]
\includegraphics[width=0.8\textwidth]{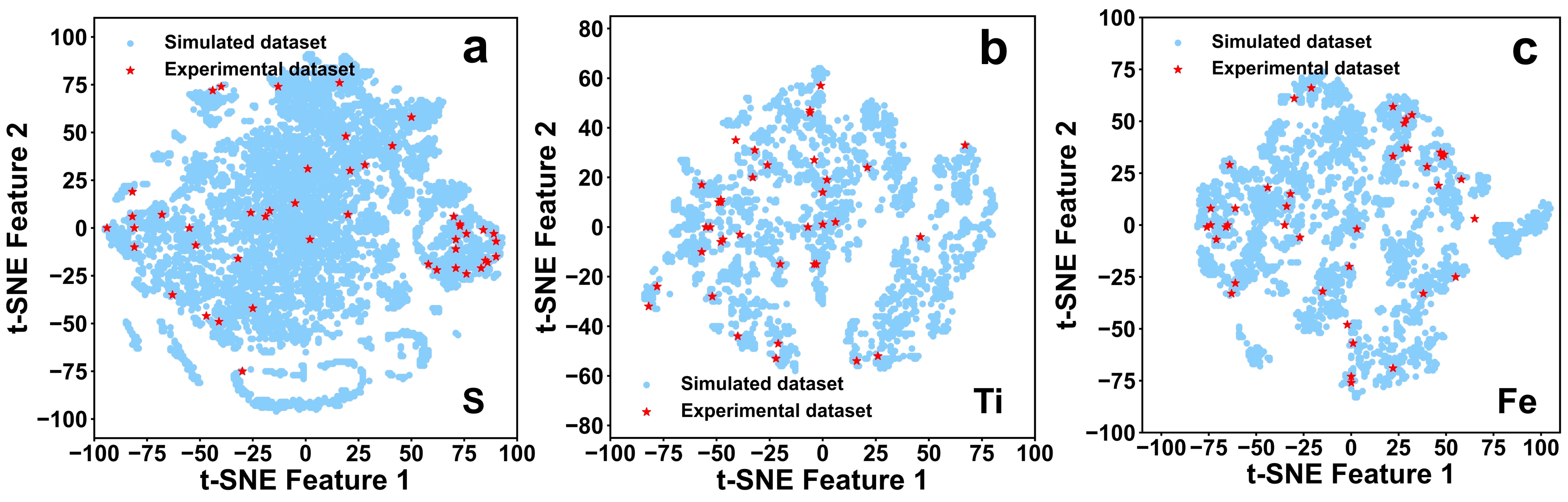}
\caption{\label{fig:distribution} Distribution of the samples from the simulated dataset and experimental dataset: (a) S K edge XANES, (b) Ti K edge XANES, (c) Fe K edge XANES. Here, the samples are represented by their spectra predicted by CGXAS\_Uni and reduced into 2 features by t-SNE. The blue dot represents the samples from the simulated dataset, and the red star represents the samples from the experimental dataset.}
\end{figure*}

\subsection{Transfer learning of CGXAS\_Uni for Experiment-calibrated XANES Prediction}
Since the CGXAS\_Uni model is trained on simulated XANES dataset and there exists difference between the simulated and experimental data, it is not surprising that we also see difference between the predicted XANES and the experiments. Therefore, it is important to introduce experimental data to calibrate the model. However, the available experimental XANES data is very limited compared with that of the simulated XANES, making it hard to train a model with high accuracy using the experimental dataset alone. To address this problem, we explore the possibility of applying transfer learning to calibrate the CGXAS\_Uni model with a small experimental dataset in this section. Here, three experimental datasets containing 48 S K edge XANES spectra, 40 Ti K edge XANES and 45 Fe K edge XANES are used in the transfer learning. The corresponded calibrated models are denoted as CGXAS\_Exp\_S, CGXAS\_Exp\_Ti, and CGXAS\_Exp\_Fe.  

The samples distribution of these 3 experimental datasets is firstly analyzed and compared with the corresponded element specific simulated datasets respectively. To illustrate the sample distribution in the simulated dataset according to the spectrum features, their predicted spectra by CGXAS\_Uni are transformed by t-SNE into two t-SNE features and plotted in the 2D space (blue dots in Fig.\ref{fig:distribution}). To evaluate the samples diversity of experimental datasets, the simulated spectra of the samples in experimental datasets predicted by CGXAS\_Uni model are marked with red stars in the spectrum space. It can be seen that the samples from experimental datasets have a wide distribution among the samples from the large simulated dataset. 

\begin{figure*}[t]
\includegraphics[width=0.8\textwidth]{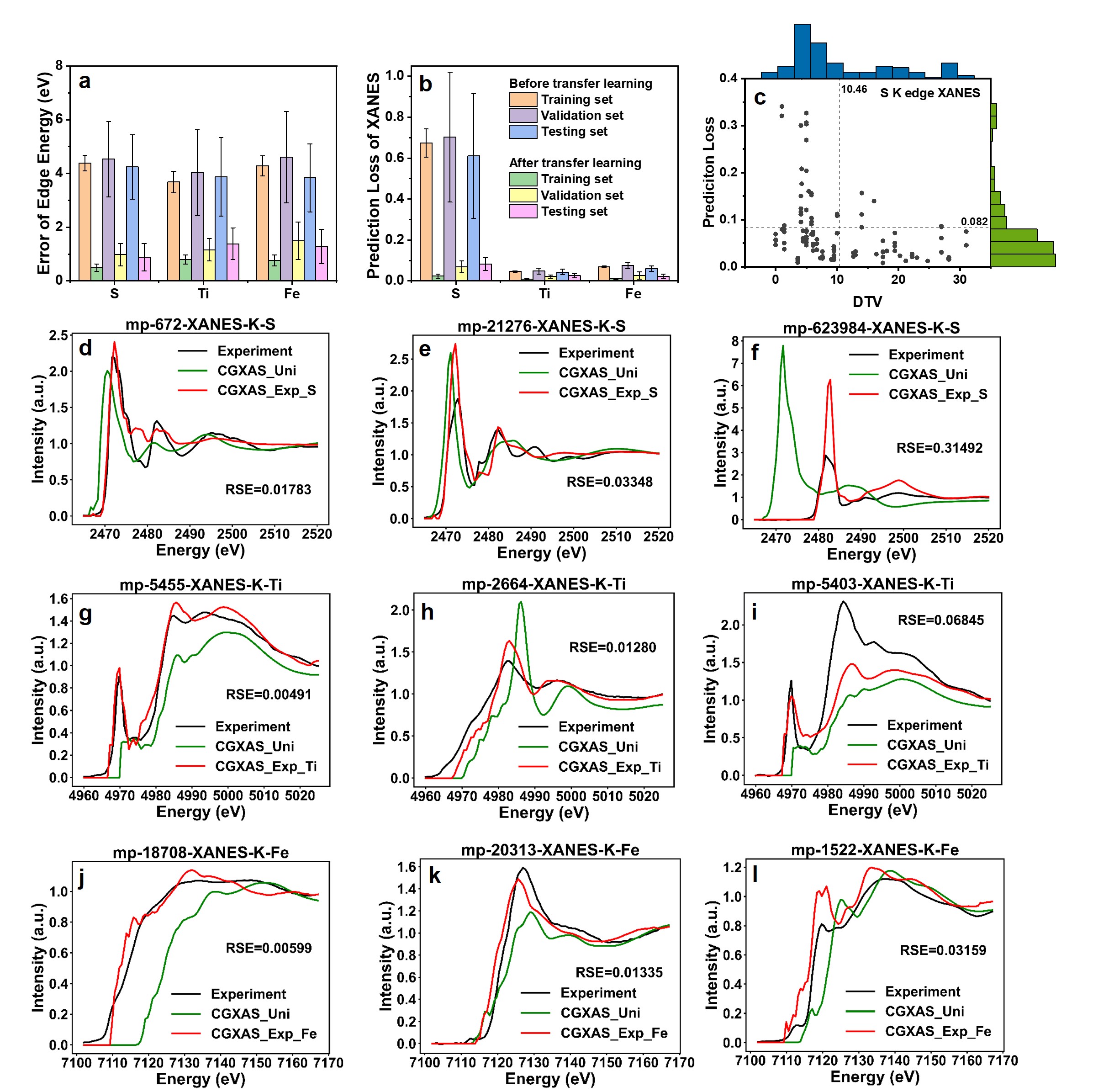}
\caption{\label{fig:transfer} Performance of the CGXAS model on the prediction of experimental XANES after transfer learning. (a) Error of edge energy compared with experimental XANES before and after transfer learning in training set, validation set, and testing set. Here, the column height and error bar represent the average value and standard deviation of the prediction losses or energy errors calculated over 20 models trained with different data separations respectively. (a) and (b) share the same legend. (b) Prediction loss of S, Ti, and Fe K edge XANES (measured by RSE) before and after transfer learning in training set, validation set, and testing set. (c) Scatter plot of the prediction loss for different samples in S K edge XANES testing sets from 20 random splits and their distance to the training and validation set (DTV). The histograms on the top and right show the distribution of these on the variances of the x axis and y axis, respectively. And the dash lines mark the average values of the variables in x and y axis. (d)-(f) Examples of S K edge XANES predicted by CGXAS\_Exp\_S for the samples in testing sets with different RSE values, together with the spectrum predicted by CGXAS\_Uni and the ground truth experimental XANES. (g)-(i) Examples of Ti K edge XANES predicted by CGXAS\_Exp\_Ti for the samples in testing sets. (j)-(l) Examples of Fe K edge XANES predicted by CGXAS\_Exp\_Fe for the samples in testing sets.}
\end{figure*}

To assess the performance of transfer learning using a small experimental dataset, the dataset is randomly separated into training, validation and testing sets, with a fixed ratio of 6:1:1, for 20 times; and each separation is used to fine-tune and test an independent model. The average value of metrics over the 20 independent models on each data split is used to evaluate the performance of transfer learning. Energy misalignment in the simulated XANES will highly influence the accuracy of some key features like absorption edge energy. To assess the ability of transfer learning to calibrate the energy scale, the absolute errors of predicted edge energy are firstly analyzed. By directly transferring the CGXAS\_Uni model on the small experimental dataset of S K edge XANES, yielding the CGXAS\_Exp\_S model, the edge energy error in testing set is reduced from 3.8 eV to 1.4 eV. The transferred model is then applied to calibrate the simulated dataset of S K edge XANES. The CGXAS\_Uni model is again finetuned on the calibrated simulated dataset and then further transferred on the experimental dataset to a new CGXAS\_Exp\_S model, resulting in a smaller edge energy error of 0.9 eV. The procedure of simulation data calibration and the multistep transfer learning can be repeated with the new CGXAS\_Exp\_S model, forming a closed cycle. Here, this procedure is cycled for 10 times, and the edge energy error is convergent to around 0.9 eV after 3 cycles. The CGXAS\_Exp\_S models from the cycle 6 with the lowest edge energy error is selected for further analysis in this work. The experiment-calibrated models for Ti (CGXAS\_Exp\_Ti) and Fe (CGXAS\_Exp\_Fe) are also trained in the same way. The edge energy error of CGXAS\_Exp\_S, CGXAS\_Exp\_Ti, and CGXAS\_Exp\_Fe in training, validation, and testing sets are shown in Fig.\ref{fig:transfer}(a). Before transfer learning, the average errors of edge energy for the XANES of these three elements are around 4 eV. The standard deviation of error before transfer learning in the validation set and testing set are much larger than that in training set, which is due to the smaller sample numbers in these two sets. After the transfer learning, the average errors of edge energy are largely reduced to around 0.8 eV by 80\% for S K edge XANES. For Ti and Fe, the errors are reduced to around 1.1 eV. The standard deviation of errors is also reduced after transfer learning. The results show that the transfer learning can effectively reduce the energy misalignment for different containing materials.

The average prediction losses in training, validation, and testing sets before and after transfer learning over the 20 models are shown in Fig.\ref{fig:transfer}(b), separately for CGXAS\_Exp\_S, CGXAS\_Exp\_Ti and CGXAS\_Exp\_Fe, where the error bar represents the standard deviation of loss. For CGXAS\_Exp\_S, the prediction losses before transfer learning are around 0.7, indicating that the simulated XANES has a large systematic error compared with the experiment measurement. The edge and peak of many predicted spectra severely misalign with experimental S K edge XANES spectra, especially for those with absorption edge appearing around 2480 eV. After the transfer learning, the average prediction losses are reduced by about 90\% to around 0.07, showing the good performance of CGXAS\_Exp\_S. For CGXAS\_Exp\_Ti and CGXAS\_Exp\_Fe, the prediction losses before transfer learning are around 0.05 and 0.07 respectively. These losses are much smaller than that of CGXAS\_S\_Exp, because the peak intensity of Ti and Fe K edge XANES is not as high as S K edge XANES in general, making the misalignment of peak yield smaller value of RSE than S K edge XANES. After transfer learning, the prediction losses are still reduced to around 0.02 for both Ti and Fe K edge XANES, showing an accordant trend with CGXAS\_Exp\_S.

Although the experimental datasets have a wide distribution in the simulated spectrum space, there are still some regions without experimental samples. Without available experimental spectrum, it’s hard to directly evaluate the calibration performance of transfer learning in these regions. In order to estimate the generalization ability of our proposed method, we collect the prediction loss of each sample from testing sets in all 20 splits, to see whether the similarity of testing sample with the corresponded training and validation sets will influence the calibration performance. Here, the similarity is measured by the minimum distance of the testing sample to all the training and validation samples (DTV) in the simulated spectrum space (Fig.\ref{fig:distribution}), which is defined in Eq.\ref{eq:DTV} in Method section. A smaller DTV value means that the testing sample has larger similarity with the training and validation sets. By this analysis, the performance of the calibrated models in the regions closed and far from the regions with experimental training data can be estimated, which can be used to indirectly assess the generalization ability of the model. The prediction losses measured by RSE of the test samples together with their DTV are shown in Fig.\ref{fig:transfer}(c) for the dataset of S. The blue and green histograms show the general distribution of DTV and prediction loss of these samples with dash line marking the average value. The DTV value in all the three cases have a wide distribution from 0 to 30. For the testing sets of S K edge XANES, comparing the samples of low (0-10), medium (10-20), and high (20-30) DTV values, their prediction losses all presents a similar distribution as the general distribution. The results show that the models have consistent performance distribution for samples with different similarities compared with training samples, implying a good generalization ability.

To further illustrate the effect of transfer learning on experimental-calibrated XANES prediction, Fig.\ref{fig:transfer}(d)-(l) show the XANES spectra predicted by CGXAS\_Exp\_S, CGXAS\_Exp\_Ti and CGXAS\_Exp\_Fe with different RSE values, as well as the corresponded spectra predicted by CGXAS\_Uni and experimental spectra. Before the transfer learning, most of the predicted XANES spectrum shows a similar shape as the experimental spectrum, but matches poorly in the absorption edge and peak positions. After the transfer learning even on a small experimental dataset, the absorption edge, the main peak positions and intensity are changed to achieve smaller difference with the experimental spectrum. For those examples with low RSE, the absorption edge and the main peak are corrected and fit well with the experimental spectrum (Fig.\ref{fig:transfer}(d), (g) and (j)). For those with medium RSE (around the average RSE), the edge position and the major shape of predicted spectra can coincide with the experimental spectra in general (Fig.\ref{fig:transfer}(e), (h) and (k)). The errors mainly come from the slight misalignment of peaks, missing of some fine structures or peak intensity difference. From Fig.\ref{fig:transfer}(c), the majority of predicted spectra are in the region of low and medium RSE. Some spectra have relatively high RSE but their edge and peak positions are still well corrected (Fig.\ref{fig:transfer}(f) and (i)), where the high RSE mainly come from the large difference in main peak intensity. For some pre-edge structures or post edge peaks with lower intensity, the calibrated models may still yield large error, which requires further optimization in future study. In general, for all the three cases of S, Ti, and Fe K edge XANES, the predicted XANES achieve better accordance with experimental results after transfer learning with a small experimental dataset, showing the good feasibility of our proposed method. 

\section{Discussion}
Different ways to achieve universal XANES prediction are compared in this work. Directly training one CGXAS\_Uni model on a universal XANES dataset shows better performance for XANES prediction of almost all the elements than training a CGXAS\_Spec\_X model on an element-specific XANES subset. Since the element-specific subset used in CGXAS\_Spec\_X comes completely from the universal dataset for a specific element X, the enhanced XANES prediction performance of the CGXAS\_Uni model must be attributed to other subsets of non-X elements in the universal dataset. Although the XANES spectra of different elements are significantly different regarding their energy ranges, the underlying physical rules for X-ray absorption and electron excitation are similar. Training model with the universal XANES dataset enables the neural network to learn the general underlying structure-XANES relationship regardless of the energy range for each element, thus enhancing the overall prediction accuracy for all elements. The fine-tuned CGXAS\_FT\_X model on the element-specific dataset has little influence on the prediction accuracy, implying that the knowledge of each element has already been well learned in the CGXAS\_Uni model. In the previous work for 3d transition metal XANES prediction, Kharel et al. came out with an opposite result: the element specific model outperforms the universal model\cite{RN365}. The cause the opposite results can be understood by the different structure features representations adopted in these two works. In this work, a GNN encoder is trained to process the structure information in the crystal graph, where a universal dataset containing more diverse structures is helpful to achieve better structure representations. In Kharel’s work, the universal dataset has no advantage in achieving better structure representations since a pretrained model is used for structure features encoding. On the contrary, the XANES prediction in universal model is more complex than that in element specific model, therefore yielding an opposite result. But pretraining on universal dataset can still reduce the overfitting risk for the finetuned universal model, therefore their finetuned universal models has better performance than the element specific models in general.

Through t-SNE analysis of the latent site vectors, the ability of CGXAS to extract the essential structural information related to XANES is verified, including absorption element, and neighboring atom type. The interpretability study and the excellent universality discussed earlier show that the XANES prediction of CGXAS\_Uni is based on learning underlying physics rather than remembering the training data. The XANES prediction process of CGXAS can be qualitatively interpreted combining the XANES theory and the network architecture shown in Fig.\ref{fig:workflow}. In the initial crystal graph, each node only contains the information of the corresponding atom and is connected with its neighboring atoms by edges. During the processing of GNN, the information of neighboring atoms is gathered into the node through the edge connections. Therefore, the site vector on each node in the outputted graph contains the information of the scattering paths between the absorption atom and its local atomic environment. Then, MLP operates the latent site vector to predict the absorption coefficient at each energy grid. 

By transfer learning on a small size experimental XANES dataset, the experimental-calibrated model can reduce the prediction loss and edge energy error compared with experimental XANES, respectively. The transfer learning strategy is helpful to break the dilemma between data fidelity and quantity. Pretrained on a large amount of low-fidelity simulated XANES enables CGXAS\_Uni model to extract essential structure information from the given crystal graph. Then, the network parameters of MLP are adjusted using a small amount of high-fidelity experimental data in the transfer learning process, which can be defined as an experimental calibration of the prediction model. Although S, Ti and Fe elements are used as examples of our transfer learning, the pretrained CGXAS\_Uni model can be easily extended to predict the experimental XANES of any other element with a similar transfer learning method. 

It should be noted that the three experiment-calibrated models are developed for the purpose of proof-of-principles in this work. There some still some problems need to be solved before they can be put into practical use. For the pre-edge structure and some finer peaks in the high energy region, the experiment-calibrated models may give relatively large errors in the prediction. And there are also a few spectra with large discrepancy with experimental spectra. In future study, the universal, efficient, and experiment-based prediction of XANES can be further enhanced by obtaining more high-fidelity XANES data and developing more powerful AI algorithms, enabling the scientist to gain deeper insights into more complex scientific problems using XAS techniques.

\section{Method}
\subsection{Node and edge features representation}
An atom in the crystal is represented by a node in the corresponding crystal graph. The atom features used in the CGXAS include the atomic number, period of atom, and the numbers of valence electrons in the s, p, d, and f orbitals, which are represented by their one-hot codes with the lengths of 101, 7, 3, 7 ,11, 15 separately (shown in Fig.\ref{fig:model}(a)). These one-hot codes are concatenated as the node vector with the length of 146. 

In the crystal graph, one atom is connected with the neighboring atoms within the cutoff radius of 8 Å through a directional edge. And the maximum number of neighbors per atom is 12 in this work. To keep the rotational invariance and encode the geometry information, the Gaussian basis set and the plane wave basis set are adopted as the edge features\cite{RN158}.

\subsection{Model architecture of CGXAS}
\begin{figure*}[ht]
\includegraphics[width=0.8\textwidth]{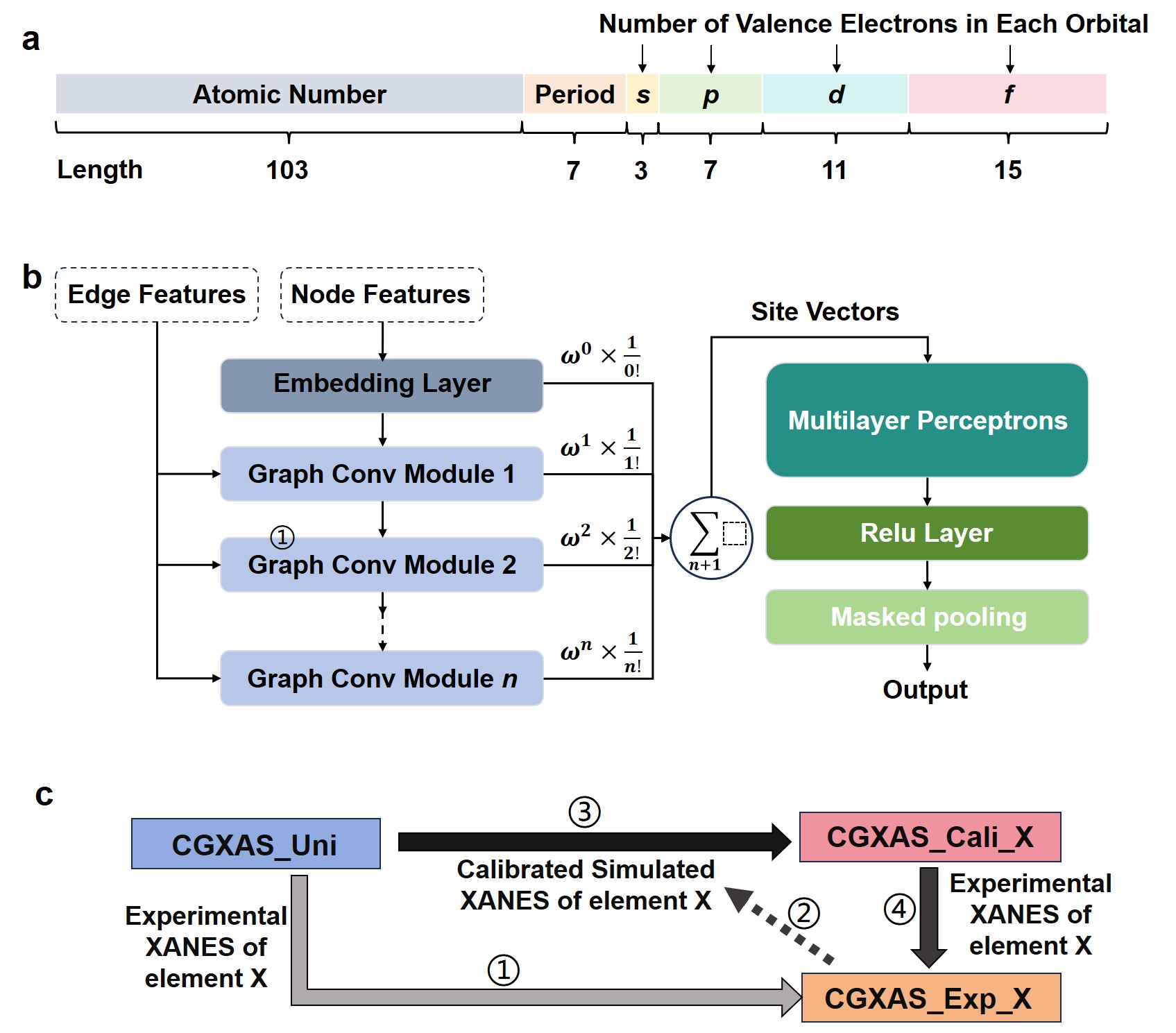}
\caption{\label{fig:model} Crystal graph representation and detailed architecture of CGXAS. (a) Node features of crystal graph. (b) Model architecture of CGXAS. (c) Flow chart of transfer learning.}
\end{figure*}

The model architecture of CGXAS is shown in Fig.\ref{fig:model}(b). The node features of the input graph are firstly embedded by an embedding layer. Then, the information of neighboring nodes is gathered to each node through the edge in the following graph convolution modules. Here, one graph convolution module is composed of a gated convolution layer of geo-CGNN\cite{RN158}, a normalization layer and a one layer perceptron. The updated node features of each graph convolution module $\omega^{i}$ are multiplied by $\frac{1}{i!}$ and summed over as the site vectors. The above modules together compose the GNN module in Fig.1. Next, the site vectors are processed by the multilayer perceptron and a Relu layer to generate the intensity of XANES of each absorption site. The activation function of the multilayer perceptron is Silu in this work. And the Relu layer is designed to avoid the prediction of negative intensity. Finally, the mask pooling layer will average the XANES over the unmasked sites, which is used to obtain the XANES of target site or specific element. The depths of gated convolution layers and multilayers perceptron are both five. The widths of input node features, hidden layer features and output spectrum are 146, 256 and 100 separately. 

\subsection{XANES dataset}
The dataset of simulated K edge XANES spectra used in this work covers the elements from the second to fifth period except the rare gas elements. The crystal structures of materials are derived from the Materials Project\cite{RN408,RN409} using the pymatgen\cite{RN289} and the XANES are obtained from an opensource XAS dataset generated through FEFF9\cite{RN219}. Each XANES spectrum is matched with a distinct atom site in a material and is indexed by a unique id in the following format: “mp id - site index - spectrum type - edge type - absorption element”. For example, “mp-561304-56-XANES-K-S” represents the S K edge XANES of the 56th site of material “mp-561304” in Materials Project. For all the XANES spectra of the same element, a unified absorption energy range that can cover the pre-edge and near edge regions is chosen. And the XANES intensities at 100 uniform energy grid points in the corresponding element energy range are defined as the output of CGXAS. By fixing the absorption energy range of different elements, the multiple elements XANES can be predicted in the same neural network. 

The experimental data used in transfer learning contains S K edge XANES of 48 different materials, Ti K edge XANES of 40 different materials, and Fe K edge XANES of 45 different materials. The spectra are obtained from beamline BL16U1 of the Shanghai Synchrotron Radiation Facility (SSRF), beamline 4B7A of Beijing Synchrotron Radiation Facility (BSRF), the XASLIB database (\url{https://xaslib.xrayabsorption.org/}), ESRF ID21 Sulfur XANES spectra database(\url{https://www.esrf.fr/home/UsersAndScience/Experiments/XNP/ID21/php.html}), MDR XAFS DB\cite{RN404}, and several literatures\cite{RN122,RN293,RN294,RN295,RN297,RN298,RN350,RN331,RN360,RN361,RN363,RN357,RN396,RN321,RN402,RN397,RN400,RN398,RN401,RN399}.

\subsection{Training of CGXAS with simulated XANES}
The simulated XANES dataset is divided into training set, validation set and testing set by the ratio of 6:2:2. The relative square error (RSE) between the actual spectrum {$\mu_{i}^{a}$ } and predicted spectrum {$\mu_{i}^{p}$ } is used as the loss metric in the training process of CGXAS\_Uni and CGXAS\_Spec\_X:
\begin{equation}
\label{eq:RSE}
    \text{RSE}=\frac{\sum_{i=1}^{100}(\mu_{i}^{a}-\mu_{i}^{p} )^{2}}{\sum_{i=1}^{100}(\mu_{i}^{a})^{2}},
\end{equation}

Adam optimizer and an initial learning rate of 0.0001 are chosen to optimize the parameters of neural network. The maximum number of epochs of training is 100. 
In the fine-tuning process of CGXAS\_FT\_X, only the parameters in the MLP are optimized while other layers in the network are unchanged. The loss metric and the hyperparameters of fine-tuning of CGXAS\_FT\_X are the same as the training process of CGXAS\_Uni and CGXAS\_Spec\_X. 
Pearson correlation coefficient (PCC) is also used to analyze the prediction accuracy of XANES in the work. The PCC between {$\mu_i^a$ } and {$\mu_i^p$ } is defined as:
\begin{equation}
\label{eq:PCC}
    \text{PCC}=\frac{\sum_{i=1}^{100}(\mu_{i}^{a}-\bar{\mu_{i}^{a}})^{2}(\mu_{i}^{p}-\bar{\mu_{i}^{p}})^{2}}{\sqrt{\sum_{i=1}^{100}(\mu_{i}^{a}-\bar{\mu_{i}^{a}})^{2}}\sqrt{\sum_{i=1}^{100}(\mu_{i}^{p}-\bar{\mu_{i}^{p}})^{2}}},
\end{equation}
where$\bar{\mu_{i}^{a}}=\frac{1}{100}\sum_{i=1}^{100}\mu_{i}^{a}$ and $\bar{\mu_{i}^{p}}=\frac{1}{100}\sum_{i=1}^{100}\mu_{i}^{p}$. A PCC value closer to 1 means that the predicted spectrum has higher accuracy. 

\subsection{Transfer learning using experimental XANES}
The transfer learning process (shown in Fig.\ref{fig:model}(c)) can be divided into the following step: 1. The experimental dataset is firstly used to finetune the CGXAS\_Uni model to a initial transferred model CGXAS\_Exp\_X. 2. The CGXAS\_Exp\_X model is applied to calibrate the energy of simulated dataset. 3. The calibrated simulated dataset is then used to finetune the CGXAS\_Uni model to the CGXAS\_Cali\_X model. 4. CGXAS\_Cali\_X model is further finetuned on the experimental dataset to the new CGXAS\_Exp\_X model. 5. Repeat step 2-4 for 10 times and select the CGXAS\_Exp\_X model with the lowest edge energy error in testing set. To avoid over fitting, only the parameters of MLP are optimized layer by layer while the other parts of network remain the same as the pretrained model. The loss metric used in transfer learning is RSE. And the size of training, validation and testing set have a ratio of 6:1:1. The Adam optimizer and an initial learning rate of 0.0001 are chosen to optimize the layer of MLP. The maximum number of training epochs of each layer is 30.

The similarity of testing set compared with training and validation is measured by the distance of testing sample to the training and validation set (DTV): 
\begin{equation}
\label{eq:DTV}
    \text{DTV}=\min\{\sqrt{(f_{1,i}-f_{1,t})^2+(f_{2,i}-f_{2,t})^2}\}
\end{equation}
where $f_{1,i}$ and $f_{2,i}$ are the two t-SNE features of the $i^{\text{th}}$ sample in training and validation set, n is the number of samples in training and validation set, and $f_{1,t}$ and $f_{2,t}$ are the two t-SNE features of the sample in testing set.

\section{Acknowledgments}
This work was supported by the National Key R\&D Program of China (2022YFA1203400) and the National Natural Science Foundation of China (W2441009). The authors thank the 4B7A Beamline of Beijing Synchrotron Radiation Facility (\url{https://cstr.cn/31109.02.BSRF.4B7A}), BL16U Beamline of Shanghai Synchrotron Radiation Facility (\url{https://cstr.cn/31124.02.SSRF.BL16U1}) and Experiment Assist System of Shanghai Synchrotron Radiation Facility (\url{https://cstr.cn/31124.02.SSRF.LAB}) for providing technical support and assistance in XANES data collection.

\bibliographystyle{apsrev4-2}
\bibliography{references.bib}

\end{document}